\documentclass[11pt,a4paper]{article}

\usepackage{jheppub}

\usepackage{enumitem,nccmath,pdflscape,subcaption,xspace}

\title{\boldmath Master integrals for splitting functions from differential equations in QCD}

\author{Oleksandr~Gituliar}
\affiliation{Instytut Fizyki J\c{a}drowej, Polska Akademia Nauk, ul.~Radzikowskiego 152, 31-342, Krak\'ow, Poland}
\emailAdd{oleksandr.gituliar@ifj.edu.pl}

\preprint{{IFJPAN-IV-2015-21}}

\DeclareMathOperator{\HPL}{\mathrm{H}}

\DeclareMathOperator{\Order}{\mathcal{O}}

\def\Fa{{\cal F}_A}
\def\Fl{{\cal F}_L}
\def\Ft{{\cal F}_T}

\newcommand{\D}{\mathrm{d}}
\newcommand{\Dps}[1]{\D\mathrm{PS}(#1)}
\newcommand{\FIRE}{\texttt{FIRE}\xspace}
\newcommand{\FORM}{\texttt{FORM}\xspace}

\newcommand{\LiteRed}{\texttt{LiteRed}\xspace}

\newcommand{\QGRAF}{\texttt{QGRAF}\xspace}
\newcommand{\Reduze}{\texttt{Reduze}\xspace}
\newcommand{\as}{\alpha_s}
\newcommand{\eps}{\epsilon}
\newcommand{\sd}[2]{\,#1\!\cdot\!#2}
\renewcommand{\eqref}[1]{eq.~(\ref{#1})}

\abstract{%
A method for calculating phase-space master integrals for the decay process \mbox{$1 \to n$} massless partons in QCD using integration-by-parts and differential equations techniques is discussed.
The method is based on the appropriate choice of the basis for master integrals which leads to significant simplification of differential equations.
We describe an algorithm how to construct the desirable basis, so that the resulting system of differential equations can be recursively solved in terms of (G)HPLs as a series in the dimensional regulator $\eps$ to any order. 
We demonstrate its power by calculating master integrals for the NLO time-like splitting functions and discuss future applications of the proposed method at the NNLO precision.
}


\begin{document} 

\maketitle
\flushbottom

\section{Introduction}

After the recent success of the LHC Run I experiment, and discovery of the Higgs boson in particular, the LHC Run II is pushing the limits of higher-order calculations in QCD even further than ever.
In that context, analytical calculations play a crucial role as a background for the numerical methods and phenomenological analyses in QCD.

In this paper, we focus on the analytical calculation of phase-space master integrals for $1 \to n$ decay processes with massless particles in the final state.
This type of decays within the electron-positron annihilation reactions have given us much information about the properties of quarks and gluons and the nature of their interactions as described by QCD.
Moreover, they will play an outstanding role for the further precision studies of QCD at upcoming $e^+ e^-$ colliders at even higher energies.
The classical example is jet production in $e^+e^-$ annihilation, which can be used to extract values of the strong coupling constant $\alpha_s$ from the three-jet rate and related event shape observables.
In the past decade, next-to-next-to-leading order (NNLO), i.e., $\Order(\as^3)$, contributions to the three-jet rate from the process $\gamma^* \to 3 \text{ partons}$ in $e^+ e^-$ annihilation were calculated \cite{FGK89,BDK97,GGGKR01,MUW02}.
Further improvements to this calculations at N$^3$LO inevitably require analytical expression for the integrals we consider in this work.
For example, three-loop splitting functions are a must-have piece for numerical calculations of N$^3$LO contributions to the three-jet rate from the $\gamma^* \to$ 6 partons process.
Splitting functions for the initial-state radiation, i.e., space-like, are known exactly at NNLO \mbox{\cite{MVV04a,MVV04b}}.
In contrast, for the final-state radiation, i.e., time-like, they are known at NNLO only approximately \cite{MMV06,MV08,AMV11}.
Despite the fact that those uncertainties are numerically irrelevant for phenomenological applications, e.g. for the evolution of fragmentation functions \cite{ARS15}. The exact result are still needed, as mentioned before, for performing numerical integration in various subtraction schemes when the need to integrate local counter terms arise \cite{GGG05,STD06,Cza10}.

At the same time, a huge progress has been made in development of tools and methods for higher-order calculations in the field theory and perturbative QCD in particular.
Integration-by-parts (IBP) reduction of Feynman integrals~\cite{CT81,Tka81} together with differential equations for master integrals~\cite{Smi04} proved, by state-of-the-art calculations, e.g. \cite{GMTW14,MSZ14,ADDHM15,AHHHKS15,BBDMS15}, to form a powerful framework for calculating high-order Feynman diagrams.
Despite that it is usually applied for virtual integrals at the level of amplitudes, this approach can be used with the same success for analytical calculation of real phase-space integrals at the level of matrix elements, where standard approaches are usually applied, i.e., to parametrize a phase space explicitly and proceed accordingly with Feynman parameters integration and similar methods~\cite{RN96,GGH03}, or alternatively to work in the Mellin space with recursion relations \cite{MV99} or a system of difference equations \cite{MM06}.

During the past few years, the method of differential equations became very popular due to the fact that a good choice of the basis for master integrals leads to significant simplifications of the differential equations \cite{Henn13}.
Although, in general, finding an appropriate basis is not easy, the approach based on the Moser algorithm \cite{Mos60} was discussed in~\cite{Henn14}, which allows to reduce the system in one singular point, but not globally.
A global extension of the Moser algorithm, which shows how to adjust the transformations in such a way that they do not spoil behavior in any other point was presented in \cite{Lee14}.
It allows for systematic simplification of differential equations.
Unfortunately, there is still no computer implementation of these methods available for the public use, which is very desirable to automatize the process.

In this work we propose an alternative method for calculating phase-space master integrals from differential equations and show how to fix boundary conditions.
The algorithm is self-consistent, in sense that if all the prerequisites are fulfilled the proof is reduced simply to verifying that proposed solutions satisfy initial equations.
The main advantage of the proposed approach is that it is relatively simple, can be easily implemented as computer code, and at the same time gives a complete solution for masters to any power in $\eps$.
Although it may be not as general as other methods, it can be successfully applied for calculating splitting functions, but is not limited only to that case.
As an example of practical use, we perform a detailed calculation of the master integrals for the NLO contribution to time-like splitting functions and discuss possible extensions to NNLO accuracy.

The paper is organized as follows: in Section \ref{sec:2} we introduce the notation and show how to calculate splitting functions from the $e^+e^-$ annihilation process.
In Section \ref{sec:deq} we formulate a solution for the system of differential equations for phase-space master integrals of the topology $1\to n$ derived from IBP reduction rules in $x$-space.
In Section \ref{sec:4} we calculate master integrals for the NLO splitting functions of $1 \to 3$ and $1 \to 4$ topologies. Finally, we discuss properties of the solutions obtained with our approach and its possible extensions to higher orders.

\section{\boldmath Splitting functions in QCD}
\label{sec:2}

Let us briefly review the main facts on splitting functions in the collinear factorization formalism of QCD, mainly for notation consistency.
For a more detailed review we refer the reader to \cite{MM06,GM15}.

Splitting functions govern the collinear evolution in hard scattering processes with hadrons in the initial (space-like) or final (time-like) state.
For processes with identified hadrons in the final state the parton-to-hadron transition is described by the parton fragmentation distributions $D_{f}^{h}(x,q^2)$, where $x$ represents the fractional momentum of the final-state parton $f$ transferred to the outgoing hadron $h$ and $q^2 \ge 0$ is a time-like hard scale.
The scale dependence of the fragmentation distributions is controlled by the so-called time-like splitting functions $P^{T}_{ba}(x)$\footnote{Further in the text we omit the superscript $T$ and assume all splitting functions to be time-like.}, and is given by
\begin{eqnarray}
\label{eq:Devol}
  {d \over d \ln q^2} \; D_{a}^{h} (x,q^2) & = &
  \int_x^1 {dz \over z} \,P^{T}_{ba} \left( z, \alpha_s(q^2) \right) D_{b}^{h} \Big(\, {x \over z},\, q^2 \Big) \; ,
\end{eqnarray}
where the summation runs over the number $n_f$ of effectively massless quark flavors and the gluon, 
$b = q_i,\bar{q}_i, g$ for $i = 1, \ldots, n_f$. 

The splitting functions $P_{ba}$ can be computed in perturbation theory in powers of the strong coupling $\alpha_s$,
\begin{eqnarray}
\label{eq:PTexp}
  P_{ba} \left( x,\alpha_s (q^2) \right) & = &
  a_s \, P_{ba}^{(0)}(x) 
  + a_s^{2}\, P_{ba}^{(1)}(x)
  + a_s^{3}\, P_{ba}^{(2)}(x) + \ldots \, ,
\end{eqnarray}
where we normalize the expansion parameter as $a_s = \alpha_s/ (4\pi)$.

As discussed at length in \cite{GM15}, splitting functions can be extracted using the mass factorization formalism from the electron-positron annihilation processes
\begin{equation}
  \label{eq:process}
  e^+ + e^- \to \gamma^*(q) \to p(k_0) + \langle\text{$n$ partons}\rangle
\end{equation}
and
\begin{equation}
  \label{eq:process-phi}
  e^+ + e^- \to \phi^*(q) \to p(k_0) + \langle\text{$n$ partons}\rangle
\end{equation}
with photon ($\gamma$) exchange and Higgs ($\phi$) boson exchange in the effective theory and a tagged parton $p = q,{\bar q},g$ with momentum $k_0$.

For the photon-exchange process~(\ref{eq:process}), following the notation in \cite{NW93}, the unpolarized differential cross-section in $m=4-2\eps$ dimensions is given by
\begin{equation}
  \frac{1}{\sigma_\text{tot}}
  \frac{\D^2 \sigma}{\D x \, \D \cos\theta}
  =
  \frac{3}{8}(1+\cos^2\theta)\, \Ft(x,\eps)
  +
  \frac{3}{4}\sin^2\theta\, \Fl(x,\eps)
  +
  \frac{3}{4}\cos\theta\,  \Fa(x,\eps),
\end{equation}
where $\theta$ denotes an angle between the beam and parton momentum $k_0$.
The scaling variable $x$ is defined as
\begin{align}
  x=\frac{2\sd{q}{k_0}}{q^2},
  &&
  q^2 = s > 0,
  &&
  0<x\le1\, .
\end{align}

\begin{figure}[h]
  \centering
  \begin{subfigure}[b]{0.3\textwidth}
    \includegraphics[width=\textwidth]{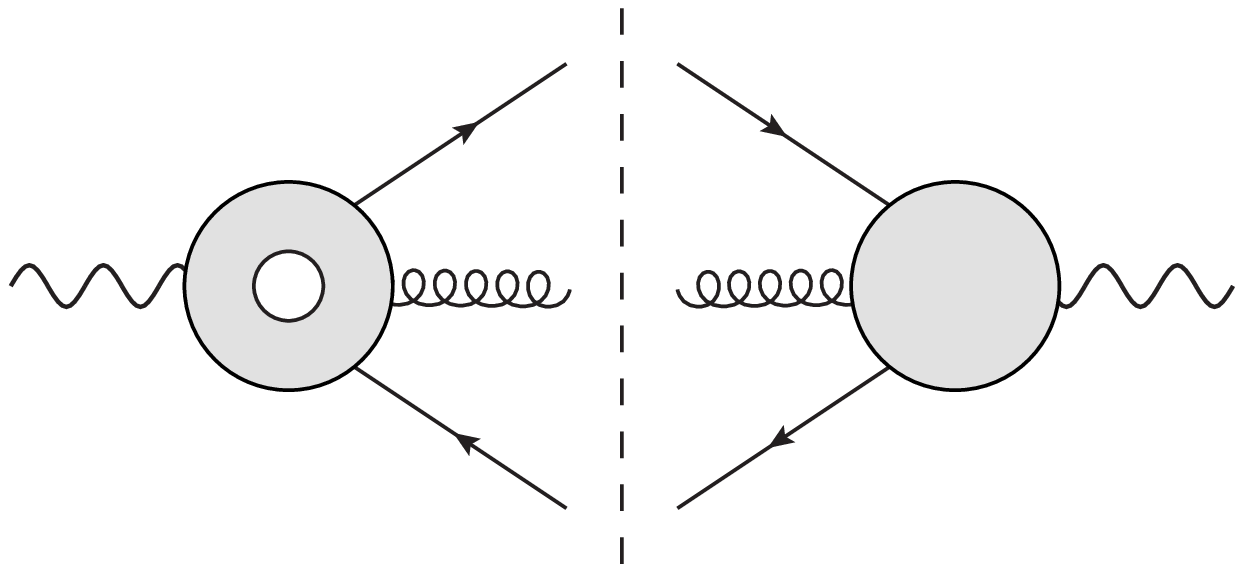}
    \caption{real-virtual $P_{qg}^{(1 \times 1)}$}
    \label{fig:amp2a}
  \end{subfigure}%
  ~
  \begin{subfigure}[b]{0.3\textwidth}
    \includegraphics[width=\textwidth]{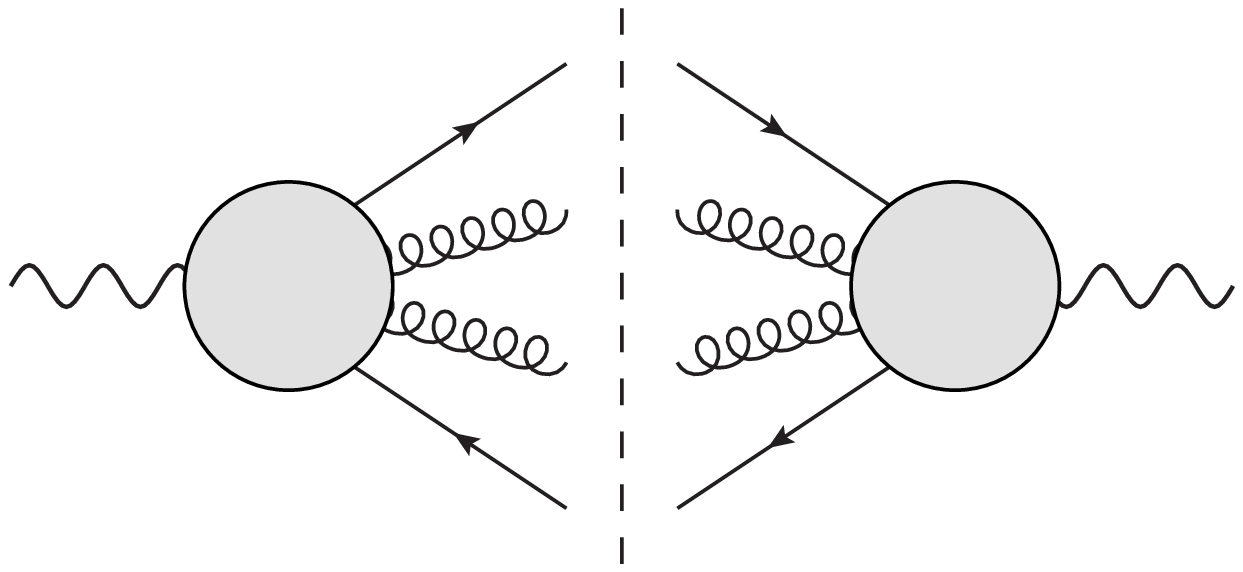}
    \caption{real-real $P_{qg}^{(2 \times 0)}$}
    \label{fig:amp2b}
  \end{subfigure}
  \caption{Contributions to the time-like splitting functions at NLO.}
  \label{fig:pqg-nlo}
\end{figure}

For the demonstration of the method for calculating master integrals described in detail in Section \ref{sec:deq}, let us consider the time-like $q \to g$ splitting function at NLO.
It can be written as
\begin{equation} \label{eq:pqg}
  P_{qg}^{(2)}(x) = \delta(1-x)\;P_{qg}^{(0 \times 2)} + P_{qg}^{(1 \times 1)}(x) + P_{qg}^{(2 \times 0)}(x),
\end{equation}
where $P_{qg}^{(n_r \times n_v)}$ denotes contribution from the diagram with $n_r$ real and $n_v$ virtual legs, as illustrated in figure~\ref{fig:pqg-nlo}.

In particular, we are interested in $1/\eps$ contribution to the transverse fragmentation function, as discussed in~\cite{GM15},
\begin{equation} \label{eq:Ft}
  {\cal F}_T^{(2)}(x,\eps)
   =
   \frac{2}{2-m} \left(
     \frac{\sd{q}{k_0}}{q^2} g^{\mu\nu} + \frac{k_0^\mu k_0^\nu}{\sd{q}{k_0}}
   \right) W_{\mu\nu}^{(2)}
,
\end{equation}
where the hadronic tensor $W_{\mu\nu}^{(2)}(x,\eps)$ for the real-virtual and real-real cases becomes
\begin{equation}
  W_{\mu\nu}^{(1\times 1)}(x,\eps) = \frac{x^{m-3}}{4\pi} \int \D \mathrm{PS}(2) \, \D l \; M_\mu (3) \: M_\nu^* (3)
\end{equation}
and
\begin{equation}
  W_{\mu\nu}^{(2\times 0)}(x,\eps) = \frac{x^{m-3}}{4\pi} \int \D \mathrm{PS}(3) \; M_\mu (4) \: M_\nu^* (4)
  ,
\end{equation}
where $M(3)$ and $M(4)$ are amplitudes for the processes depicted in figures~\ref{fig:amp2a} and \ref{fig:amp2b} respectively, $l$ is a loop momentum, and $\Dps{n}$ denotes a $n$-particle phase-space integral
\begin{equation} \label{eq:psn}
    \int \D \mathrm{PS}(n)
    =
    \int \prod_{i=0}^{n} \D^m\! k_i \; \delta^+(k_i^2) \; \delta\Big(x-\frac{2\sd{q}{k_0}}{q^2}\Big) \; \delta\Big(q-\sum_{j=0}^{n} k_j\Big)
.
\end{equation}
In the Section~\ref{sec:deq} we provide a method to calculate this kind of integrals with some detailed examples.
Below, however, we would like to present a general plan of the calculation:
\begin{enumerate}
  \item Generate amplitudes in figure~\ref{fig:pqg-nlo} with \QGRAF~\cite{Nog91} and construct from them fragmentation function $\mathcal{F}_T(x,\eps)$ using \FORM~\cite{Ver00}.
  \item Generate integration-by-parts rules for phase-space integrals with the \LiteRed package~\cite{Lee12,Lee13}.
  \item Find master integrals solving differential equations in $x$-space as described in the next section.
\end{enumerate}

\section{Master integrals from differential equations}
\label{sec:deq}

We consider a homogeneous system of differential equations, which in the most general case takes the form
\begin{equation} \label{eq:f}
  \frac{\partial f_i}{\partial x} = \sum_{j=1}^{n} a_{ij}(x,\eps) \, f_j(x,\eps),
\end{equation}
where the coefficients $a_{ij}(x,\eps)$ (or the $n\times n$ matrix $A(x,\eps)$) are known, $f_i(x,\eps)$ are unknown functions, and $\eps$ is an infinitesimally small parameter (playing the role of a dimensional regulator in $m=4-2\eps$ dimensions).

Assuming that the coefficients $a_{ij}(x,\eps)$ are rational functions of $\eps$, without loss of generality they can be written in the form
\begin{equation} \label{eq:a_eps}
  A(x,\eps) = \sum_{k=r_\eps}^\infty \eps^k \, A^{(k)}(x),
\end{equation}
where $r_\eps$ is an integer (likely negative), which we use to denote an {\em $\eps$-rank} of the matrix $A(x,\eps)$.

On the other hand, we restrict the matrix $A(x,\eps)$ to have the form
\begin{equation} \label{eq:a_x}
  A(x,\eps) = \sum_{i} \frac{A_i(x,\eps)}{(x-x_i)^{1-p_i}},
\end{equation}
where $i$ runs over some finite set, $p_i$ is said to be the {\em Poincare rank} of $A_i(x,\eps)$ at a singular point $x_i$, and $A_i(x,\eps)$ is a regular matrix at $x=x_i$, i.e., polynomial.
Such a form is imposed exclusively for a practical reason since calculations of the splitting functions are bound to the case of $x_i \in \{-1,0,1\}$, which is exactly an alphabet for Harmonic Polylogarithms (HPLs)~\cite{RV99}.
In the case of a more complex structure of denominators in the expansion \eqref{eq:a_x} the same arguments could be extended to the more general case of Generalized Harmonic Polylogarithms (GHPLs) introduced in~\cite{AB04}, which maintain the structure and properties of HPLs~\cite{BDV10,ABS13}.

Keeping all the above considerations in mind we proceed with providing a solution for~\eqref{eq:f} as an $\eps$-series.
Taking into account a recursive definition of (G)HPLs we show that such a series can be found to any order in~$\eps$ at a low computational price.

\subsection{Solutions for $\eps$-rank $>0$} \label{sec:eps>0}

We are looking for the solution of the system \eqref{eq:f} in the form
\begin{equation}
  f_i(x,\eps) = \sum_{k=1}^\infty \eps^k f_i^{(k)}(x).
\end{equation}
Keeping in mind the expansion \eqref{eq:a_eps} it is easy to show that expansion coefficients calculated by the recursive formula
\begin{equation} \label{eq:f_sol}
  f_i^{(k)}(x) = c_i^{(k)} + \sum_{m=1}^k \int \D x \, a_{ij}^{(m)}(x) f_j^{(k-m)}(x)
\end{equation}
lead to the desired solution, where $c_i^{(k)}$ are integration constants determined from boundary conditions as described in Section~\ref{sec:boundary}.

\subsection{Solutions for $\eps$-rank $=0$}

There is no general solution for the system with $\eps\text{-rank}=0$, however for some special cases it is possible to write down such a solution.
In this paper we consider {\em weakly coupled} systems, these are systems for which $a_{ij}^{(0)}(x)$ is a triangular matrix, i.e.,
\begin{equation}
  a_{ij}^{(0)}(x) = 0, \quad\text{ for }\quad i<j.
\end{equation}
In such a case it is possible to choose a new basis so that a new system has $\eps$-rank $>0$ and can be solved using the method of Section~\ref{sec:eps>0}.
In the remaining part of this section we provide a procedure how to accomplish that task, which consists of finding such new bases that:
\begin{enumerate}[label=\roman*)]
  \item diagonal elements of $a_{ij}^{(0)}(x)$ are zero, i.e., $a_{ij}^{(0)}(x)=0$ for $i=j$; and
  \item off-diagonal elements of $a_{ij}^{(0)}(x)$ are zero, i.e., $a_{ij}^{(0)}(x)=0$ for $i>j$.
\end{enumerate}

\subsubsection{Zero-diagonal form}
\label{sec:321}

It is easy to verify that a system of differential equations for a new basis defined as
\begin{equation} \label{eq:g-f}
  g_i(x,\eps) = b_{ii}(x,\eps) f_i(x,\eps),
\end{equation}
where
\begin{equation}
  b_{ii}(x,\eps) = \exp\left(-\int \D x \; a_{ii}(x,\eps)\right),
\end{equation}
contains zeroes as diagonal elements and has a new form
\begin{equation} \label{eq:g}
  \frac{\partial g_i}{\partial x} = \sum_{j=1}^{n} \tilde{a}_{ij}(x,\eps) \, g_j(x,\eps), \quad \text{where} \quad \tilde{a}_{ij}(x,\eps) = \frac{a_{ij}(x,\eps)}{b_{jj}(x,\eps)}.
\end{equation}

\subsubsection{Zero-triangular form}
\label{sec:322}

Next, following the same strategy, we find a new basis $h_i(x,\eps)$ which leads to the zero-triangular form of the differential equations:
\begin{equation} \label{eq:h_ij}
  h_i(x,\eps) = g_i(x,\eps) + \sum_{j=1}^{i-1} b_{ij}(x,\eps) g_j(x,\eps),
\end{equation}
where
\begin{equation} \label{eq:h_b_ij}
  b_{ij}(x,\eps) = - \int \D x \bigg( \tilde{a}_{ij}^{(0)}(x) + \sum_{k=j+1}^{i-1} b_{ik}(x,\eps) \tilde{a}_{kj}^{(0)}(x) \bigg).
\end{equation}
A complete form of the new system is rather complex and it is of no practical use to write it down here. However it can be easily obtained from \eqref{eq:h_ij} after coefficients of \eqref{eq:h_b_ij} are explicitly calculated.

Let us show that such a choice indeed provides the desired zero-triangular system of equations.
Taking the derivative of \eqref{eq:h_ij}, keeping in mind \eqref{eq:g} and \eqref{eq:h_b_ij}, and neglecting higher-order term in $\eps$ we obtain
\begin{equation} \label{eq:part_h}
  \frac{\partial h_i}{\partial x}
  =
 \sum_{j=1}^{i-1}
 \bigg(
   \tilde{a}_{ij}^{(0)} g_j - \bigg( \tilde{a}_{ij}^{(0)} g_j + \sum_{k=j+1}^{i-1} b_{ik} \tilde{a}_{kj}^{(0)} g_j \bigg) +\sum_{k=1}^{j-1} b_{ij}  \tilde{a}_{jk}^{(0)} g_k
 \bigg)
 .
\end{equation}
It is easy to check, by carefully switching summation variables in one of the nested sums, that right-hand side of \eqref{eq:part_h} becomes zero.

At first sight, it may look that nested integrals in \eqref{eq:h_b_ij} are way too complicated for practical calculations.
In fact, they are very easy to compute taking into account the recursive nature of (G)HPLs, as was discussed earlier in this section.
For our examples, discussed in the next section, we have used the \texttt{HPL} package \cite{Maitre05}.

\subsection{Solutions for $\eps$-rank $<0$}
\label{sec:33}

As a rule, when one chooses a basis of master integrals as provided directly by the IBP rules generator, like \FIRE \cite{Smi08}, \Reduze \cite{MS12}, or \LiteRed \cite{Lee12,Lee13}, the system \eqref{eq:f} has a negative $\eps$-rank.
In this situation we can not proceed with the procedure described before in this section.
To overcome this issue it is usually enough to adjust $\eps^{n}$ factors in the masters, for example see \eqref{eq:v_i_1} in Appendix~\ref{app:a}.

To get a hint on how to choose $n$ we analyze Mellin moments for the corresponding masters that leads to several possibilities:
\begin{enumerate}
  \item In the presence of factors $x^{-1+a\eps}(1-x)^{-1+b\eps}$ we choose $n = r_\eps^{(1)} - 1$, where $r_\eps^{(i)}$ is an $\eps$-rank of the $i^\text{th}$ Mellin moment.
  The reason is that the logarithmic singularity in $x$ is canceled by the Mellin moment while the second one in $1-x$ introduces an additional $\eps$ pole.
  For the illustration see masters $V_6, R_7, R_8$ in the Appendices.
  \item In the presence of factors $x^{-1+a\eps}$ we choose $n = r_\eps^{(0)} - 1$.
  \item Otherwise we choose $n = r_\eps^{(0)}$.
\end{enumerate}

\subsection{Boundary conditions} \label{sec:boundary}

The final step of the method is to find integration constants $c_i^{(k)}$ emerging in \eqref{eq:f_sol}.
On the one hand, in the case of phase-space integrals we can do that by calculating Mellin moments of the solution \eqref{eq:f_sol}.
On the other hand, the same moments can be taken from the literature or directly calculated by performing integration over the entire $n$-particle phase-space, i.e.,
\begin{equation}
    \int \prod_{i=0}^{n} \D^m\! k_i \; \delta^+(k_i^2) \; \delta\Big(q-\sum_{j=0}^{n} k_j\Big).
\end{equation}
As in the case of the phase-space integrals with $x$-space projection \eqref{eq:psn}, we can generate IBP rules for the inclusive integrals as well.
That allows us to reduce the set of inclusive masters which should be calculated explicitly.

Another simplification is related to the Mellin moments, which can be extracted from the difference equations.
These equations in turn can be derived from the differential equations \eqref{eq:f}, hence only one Mellin moment needs to be computed for each inclusive master.

\section{\boldmath Master integrals for NLO splitting functions}
\label{sec:4}

Finally, we demonstrate the practical application of the method described in the previous section.
We choose to calculate two-loop contributions to the time-like splitting function $P_{qg}^{(2)}(x)$ since its three-loop contribution is still not known exactly, however it will be possible to obtain them by future extension of this example to NNLO.

We follow the plan described at the end of Section~\ref{sec:2}.
After IBP reduction, done with the help of \LiteRed, we obtain 6 real-virtual (figure~\ref{fig:1}) and 8 real-real (figure~\ref{fig:2}) masters for contributions depicted in figure~\ref{fig:pqg-nlo}.

\subsection{Real-virtual contribution}

\begin{figure}[h]
  \centering
  \begin{subfigure}[b]{0.250\textwidth}
    \includegraphics[width=\textwidth]{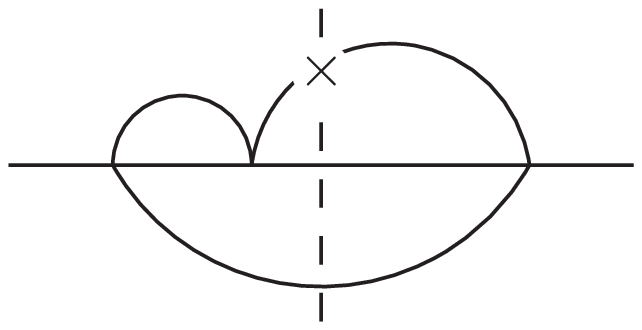}
    \caption*{$V_1 = \{1,2\}$}
  \end{subfigure}%
  ~
  \begin{subfigure}[b]{0.250\textwidth}
    \includegraphics[width=\textwidth]{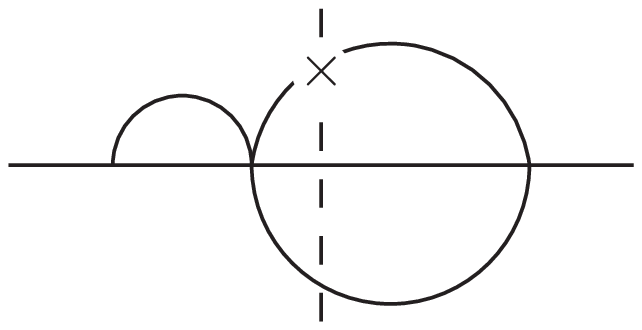}
    \caption*{$V_2 = \{1,3\}$}
  \end{subfigure}
  ~
  \begin{subfigure}[b]{0.250\textwidth}
    \includegraphics[width=\textwidth]{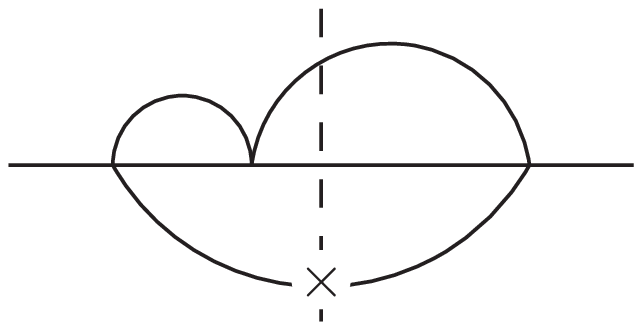}
    \caption*{$V_3 = \{1,4\}$}
  \end{subfigure}
  \vspace{4mm}

  \begin{subfigure}[b]{0.250\textwidth}
    \includegraphics[width=\textwidth]{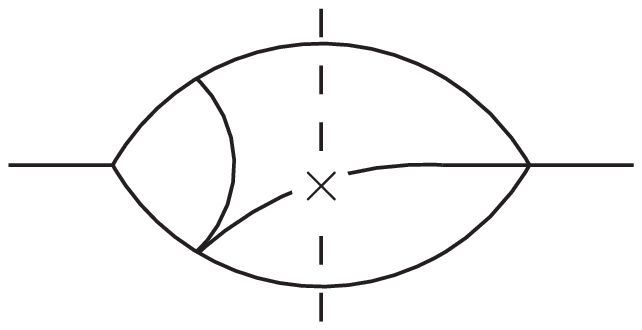}
    \caption*{$V_4 = \{1,2,3\}$}
  \end{subfigure}
  ~
  \begin{subfigure}[b]{0.250\textwidth}
    \includegraphics[width=\textwidth]{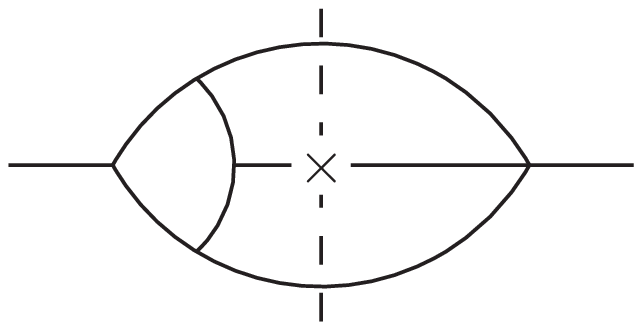}
    \caption*{$V_5 = \{1,2,3,5\}$}
  \end{subfigure}
  ~
  \begin{subfigure}[b]{0.250\textwidth}
    \includegraphics[width=\textwidth]{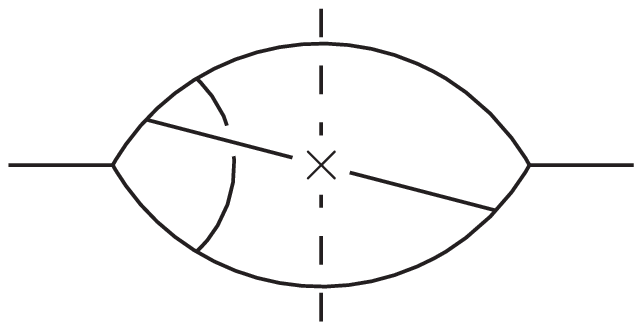}
    \caption*{$V_6 = \{1,2,3,6,7\}$}
  \end{subfigure}
  \vspace{4mm}
  \caption{Master integrals for the real-virtual NLO contribution to the time-like splitting function.}
  \label{fig:1}
\end{figure}

We define real-virtual master integrals depicted in figure~\ref{fig:1} as
\begin{equation}
  V_i(x,\eps)
  =
  \{a_1,\ldots,a_n\}
  =
  \int \D \mathrm{PS}(2) \, \D l
    \frac{1}{D_{a_1} \ldots D_{a_n}}
,
\end{equation}
where real-state integration phase-space is defined by \eqref{eq:psn}, $l$ is a loop momentum, and denominators $D_j$ are defined in \eqref{eq:Drv}.

\begin{equation}
\begin{aligned}
  D_1 & = l^2
  &
  D_2 & = (l+k_1-q)^2
  &
  D_3 & = (l-q)^2 
  &
  D_4 & = (l+k_1+k_2)^2 
  \\
  D_5 & = (l-k_2)^2
  &
  D_6 & = (l+k_1+k_2-q)^2 
  &
  D_7 & = (k_2-q)^2.
\end{aligned}
\label{eq:Drv}
\end{equation}

{\bf Step 1.} In order to obtain a system of differential equations with non-negative $\eps$-rank we choose $\eps^n$ factors as described in Section~\ref{sec:33} (see \eqref{eq:v_i_1}).
The resulting system is given by \eqref{eq:mrv}.

{\bf Step 2.} We change the basis to obtain a zero-diagonal system, as described in Section~\ref{sec:321}:
\begin{equation}
\label{eq:v_i_1}
\begin{aligned}
V_1 & = \eps x^{-1+2\eps} (1-x)^{2\eps} V_1
\\
V_2 & = \eps x^{-1+3\eps} (1-x)^\eps V_2
\\
V_3 & = \eps x^{-1+2\eps} (1-x)^\eps V_3
\\
V_4 & = \eps^2 (1-x)^{2\eps} V_4
\\
V_5 & = \eps^3 x^{1+2\eps} (1-x)^{1+2\eps} V_5
\\
V_6 & = \eps^3 x^{1+4\eps} V_6
\end{aligned}
\end{equation}

{\bf Step 3.} We make the last change of the basis in order to obtain a zero-triangular system as described in Section~\ref{sec:322}:
\begin{equation}
\begin{aligned}
V_1 & = \eps\, x^{-1+2\eps} (1-x)^{2\eps} V_1
\\
V_2 & = \eps\, x^{-1+3\eps} (1-x)^\eps V_2
\\
V_3 & = \eps\, x^{-1+2\eps} (1-x)^\eps V_3
\\
V_4 & = - \eps\, x^{-1+3\eps} (1-x)^{\eps} \HPL_{1} V_2 + \eps\, x^{-1+2\eps} (1-x)^{\eps} \HPL_{1} V_3 + \eps^2 (1-x)^{2\eps} V_4
\\
V_5 & = \eps^3 x^{1+2\eps} (1-x)^{1+2\eps} V_5
\\
V_6 & = \eps^3 x^{1+4\eps} V_6
\end{aligned}
\end{equation}

{\bf Step 4.}
We solve the resulting equations with the help of \eqref{eq:f_sol} as described in Section~\ref{sec:eps>0} and return to the initial basis.\\

{\bf Step 5.}
We find the final result by fixing boundary conditions using Mellin moments given in Appendix~\ref{app:a}.

\subsection{Real-real contribution}

\begin{figure}[h]
  \centering
  \begin{subfigure}[b]{0.200\textwidth}
    \includegraphics[width=\textwidth]{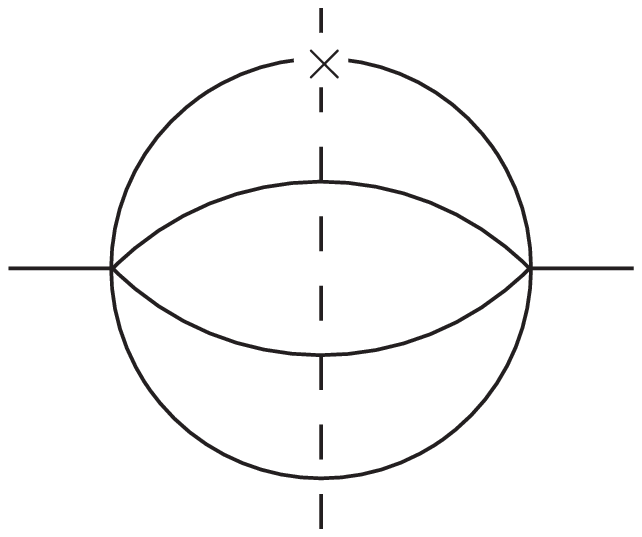}
    \caption*{$R_1 = \{\}$}
  \end{subfigure}%
  ~
  \begin{subfigure}[b]{0.200\textwidth}
    \includegraphics[width=\textwidth]{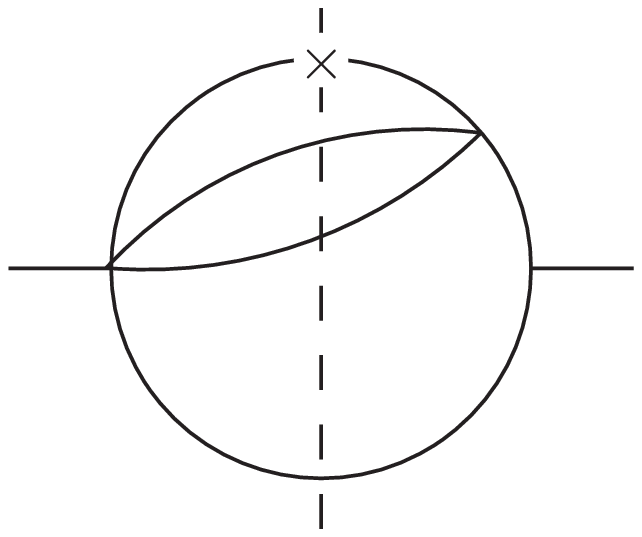}
    \caption*{$R_2 = \{2\}$}
  \end{subfigure}
  ~
  \begin{subfigure}[b]{0.200\textwidth}
    \includegraphics[width=\textwidth]{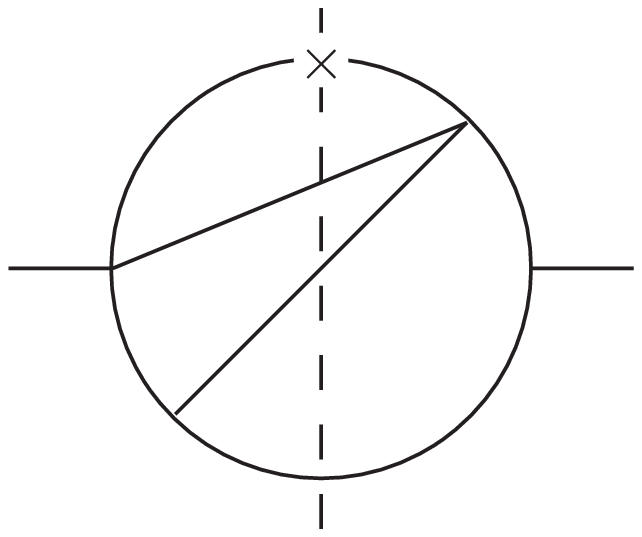}
    \caption*{$R_3 = \{3,6\}$}
  \end{subfigure}
  ~
  \begin{subfigure}[b]{0.200\textwidth}
    \includegraphics[width=\textwidth]{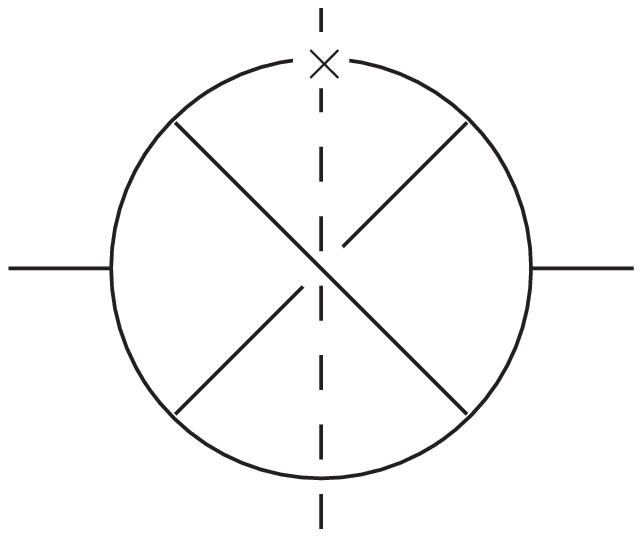}
    \caption*{$R_4 = \{4,5,6,7\}$}
  \end{subfigure}
  \vspace{4mm}

  \begin{subfigure}[b]{0.200\textwidth}
    \includegraphics[width=\textwidth]{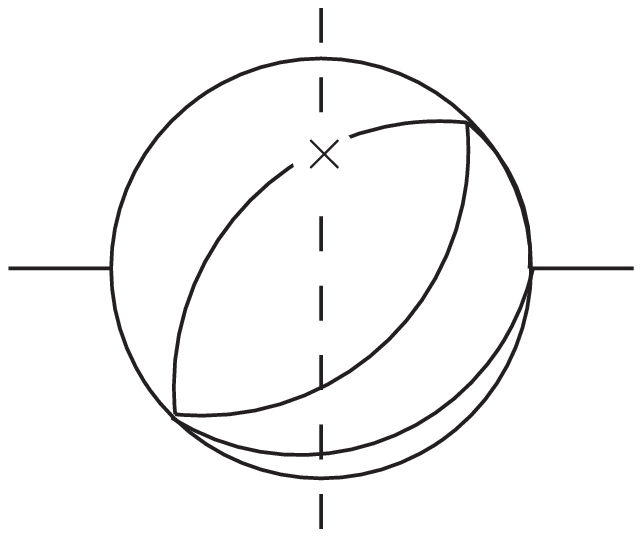}
    \caption*{$R_5 = \{1,2,3\}$}
  \end{subfigure}
  ~
  \begin{subfigure}[b]{0.200\textwidth}
    \includegraphics[width=\textwidth]{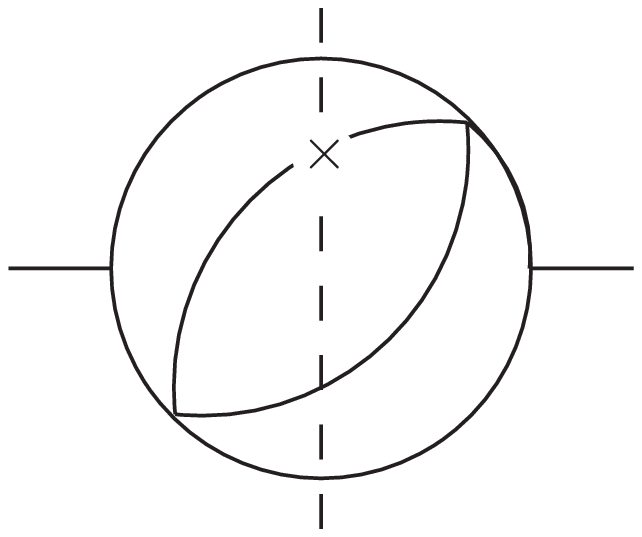}
    \caption*{$R_6 = \{2,3\}$}
  \end{subfigure}
  ~
  \begin{subfigure}[b]{0.200\textwidth}
    \includegraphics[width=\textwidth]{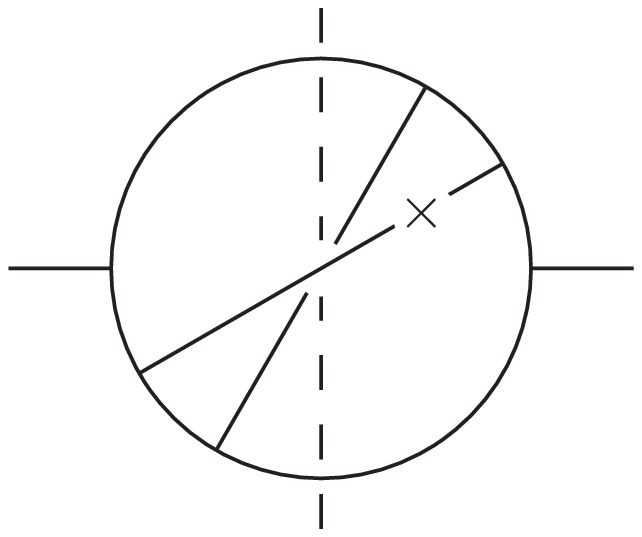}
    \caption*{$R_7 = \{2,3,6,7\}$}
  \end{subfigure}
  ~
  \begin{subfigure}[b]{0.200\textwidth}
    \includegraphics[width=\textwidth]{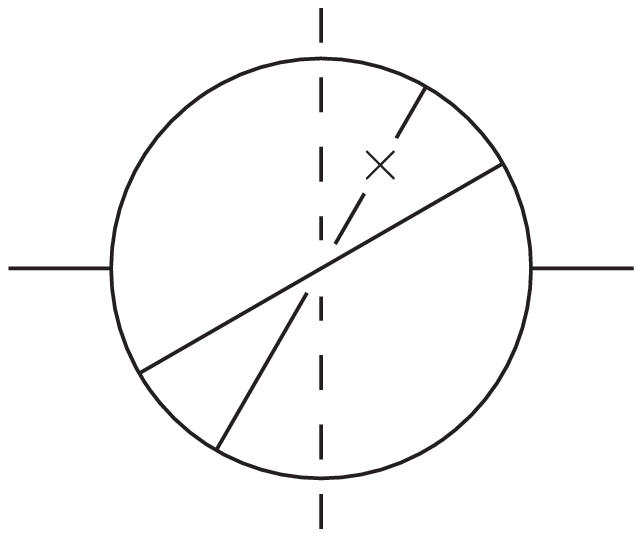}
    \caption*{$R_8 = \{2,3,4,5\}$}
  \end{subfigure}
  \vspace{4mm}
  \caption{Master integrals for the real-real NLO contributions to the time-like splitting function.}
  \label{fig:2}
\end{figure}

By analogy with the real-virtual case we proceed with the real-real contribution with final results given in Appendix~\ref{app:b}.
We define master integrals depicted in figure~\ref{fig:2} as
\begin{equation}
  R_i(x,\eps)
  =
  \{a_1,\ldots,a_n\}
  =
  \int \D \mathrm{PS}(3)
    \frac{1}{D_{a_1} \ldots D_{a_n}}
,
\end{equation}
where denominators $D_j$ are defined in \eqref{eq:Drr}.
\begin{equation}
\label{eq:Drr}
\begin{aligned}
  D_1 & = k_1^2 
  &
  D_2 & = (q-k_1)^2
  &
  D_3 & = (q-k_2)^2
  &
  D_4 & = (q-k_1-k_3)^2 
  \\
  D_5 & = (q-k_2-k_3)^2
  &
  D_6 & = (k_2+k_3)^2 
  &
  D_7 & = (k_1+k_3)^2
\end{aligned}
\end{equation}

We tune $\eps^n$ factors in order to obtain a non-negative $\eps$-rank matrix which is given by \eqref{eq:mrr}.
Corresponding powers of $\eps$ can be seen in \eqref{eq:rr-zd}.

Next, a zero-diagonal basis reads
\begin{equation}
\label{eq:rr-zd}
\begin{aligned}
R_1 & = x^{-1+2\eps} (1-x)^{-1+2\eps} R_1
\\
R_2 & = x^{-1+3\eps} R_2
\\
R_3 & = \eps (1-x)^{2\eps} R_3
\\
R_4 & = \eps^3 x^{1+2\eps} (1-x)^{1+2\eps} R_4
\\
R_5 & = \eps^2 x^{4\eps} (1-x)^{2\eps} (1+x)^{-4\eps} R_5
\\
R_6 & = \eps^2 (1+x)^{-1+6\eps} R_6
\\
R_7 & = \eps^3 x^{1+2\eps} (1-x)^{1+2\eps} R_7
\\
R_8 & = \eps^2 x^{1+2\eps} (1+x)^{1+2\eps} R_8
\end{aligned}
\end{equation}

Finally, a zero-triangular basis reads
\begin{equation}
\begin{aligned}
R_1 & = x^{-1+2\eps} (1-x)^{-1+2\eps} R_1
\\
R_2 & = x^{-1+3\eps} R_2 + 2 x^{-1+2\eps} (1-x)^{-1+2\eps} \HPL_0 R_1
\\
R_3 & = \eps (1-x)^{2\eps} R_3 - x^{-1+3\eps} \HPL_1 R_2 + 2 x^{-1+2\eps} (1-x)^{-1+2\eps} \HPL_2 R_1
\\
R_4 & = \eps^3 x^{1+2\eps} (1-x)^{1+2\eps} R_4
\\
R_5 & = \eps^2 x^{4\eps} (1-x)^{2\eps} (1+x)^{-4\eps} R_5
\\
R_6 & = 2 x^{-1+2\eps} (1-x)^{-1+2\eps} (1+x)^{-1} R_1 + 2 \eps^2 x^{4\eps} (1-x)^{2\eps} (1+x)^{-1-4\eps} R_5
    \\
    & + \eps^2 (1+x)^{-1+6\eps} R_6
\\
R_7 & = \eps^3 x^{1+2\eps} (1-x)^{1+2\eps} R_7
\\
R_8 & = 4 x^{-1+2\eps} (1 - x)^{-1+2\eps} (1 + x)^{-1} ((1+2x)\HPL_0 + (1-x) \HPL_{-1}) R_1
    \\
    & - 4\eps^2 x^{4\eps} (1-x)^{2\eps} (1+x)^{-1-4\eps} (\HPL_0 - (1-x) \HPL_{-1}) R_5
    \\
    & + \eps^2 (1+x)^{-1+6\eps}(-2\HPL_0 + 4 \HPL_{-1}) R_6 + \eps^2 x^{1+2\eps} (1+x)^{1+2\eps} R_8
\end{aligned}
\end{equation}

\begin{landscape}
\vspace*{\fill}

\begin{equation}
\label{eq:mrv}
\hspace{-8mm}
\begin{pmatrix}
 \frac{1-x - \eps (2-4x)}{x(1-x)} & 0 & 0 & 0 & 0 & 0 \\
 0 & \frac{1-x-\eps (3-4x)}{x(1-x)} & 0 & 0 & 0 & 0 \\
 0 & 0 & \frac{1-x-\eps (2-3 x)}{x(1-x)} & 0 & 0 & 0 \\
 0 & -\frac{1-5\eps+6\eps^2}{x(1-x)} & \frac{(1-2 \eps)^2}{x(1-x)} & \frac{2 \eps}{1-x} & 0 & 0 \\
 0 & \frac{\eps(3-x)\left(1-5\eps+6\eps^2\right)}{x^3(1-x)^2} & -\frac{2\eps(1-2\eps)^2}{x^2(1-x)^2} & -\frac{2 \eps^2}{x(1-x)^2} & -\frac{(1+2\eps) (1-2x)}{(1-x) x} & 0 \\
 0 & -\frac{2 \eps \left(1-5\eps+6\eps^2\right)}{x^2(1-x)} & 0 & -\frac{4 \eps^2}{x(1-x)} & 0 & -\frac{1+4\eps}{x} \\
\end{pmatrix}
\end{equation}

\vspace*{\fill}

\begin{equation}
\label{eq:mrr}
\hspace{-8mm}
\begin{pmatrix}
 \frac{(1-2\eps)(1-2x)}{x(1-x)} & 0 & 0 & 0 & 0 & 0 & 0 & 0 \\
 -\frac{2-3 \eps}{x(1-x)} & \frac{1-3 \eps}{x} & 0 & 0 & 0 & 0 & 0 & 0 \\
 0 & -\frac{1-5\eps+6\eps^2}{(1-x) x} & \frac{2 \eps}{1-x} & 0 & 0 & 0 & 0 & 0 \\
 0 & 0 & 0 & -\frac{(1+2\eps) (1-2x)}{x(1-x)} & 0 & 0 & 0 & 0 \\
 -\frac{\eps \left(2-13\eps+27\eps^2-18\eps^3\right)}{x^2(1+x)} & -\frac{\eps^2 \left(1-5\eps+6\eps^2\right)}{x^2} & 0 & 0 & -\frac{2 \eps \left(2-3x-x^2\right)}{x(1-x)(1+x)} & -\frac{2 \eps (1-6\eps)}{x (1+x)} & 0 & 0 \\
 \frac{2-13\eps+27\eps^2-18\eps^3}{x(1-x)(1+x)} & 0 & 0 & 0 & \frac{2}{1+x} & \frac{1-6\eps}{1+x} & 0 & 0 \\
 \frac{4 \eps \left(2-13\eps+27\eps^2-18\eps^3\right)}{x^3(1-x)^3(1+x)} & \frac{2 \eps^2 \left(1-5\eps+6\eps^2\right) (2-x)}{x^3(1-x)^2} & -\frac{4 \eps^3}{x(1-x)^2} & 0 & \frac{4 \eps (1+x^2)}{x^2(1-x)^2(1+x)} & \frac{2 \eps (1-6\eps)}{x^2(1-x)(1+x)} & -\frac{(1+2\eps) (1-2x)}{x(1-x)} & 0 \\
 -\frac{2 \left(2-13\eps+27\eps^2-18\eps^3\right) (1+4x+x^2)}{x^3(1-x)(1+x)^3} & -\frac{2\eps \left(1-5\eps+6\eps^2\right) (1-x)}{x^3 (1+x)^2} & 0 & 0 & \frac{4 (1+x^2)}{x^2 (1+x)^3} & \frac{2(1-6\eps)(1-x)}{x^2 (1+x)^3} & 0 & -\frac{(1+2\eps)(1+2x)}{x(1+x)} \\
\end{pmatrix}
\end{equation}

\vspace*{\fill}
\end{landscape}

In summary, with the help of the method described in Section~\ref{sec:deq} we have found the master integrals of figures \ref{fig:1} and \ref{fig:2}.
The solutions are presented in Appendix~\ref{app:a} as a partly-expanded series in the dimensional regulator $\eps$ with at least 3 leading terms, i.e., HPLs of weight 2.
That is enough for our purpose, i.e., to extract slitting functions as discussed in Section~\ref{sec:2}.
Furthermore, the higher-order $\eps$-terms of the presented solutions are easy to obtain provided the corresponding $\eps$-terms of Mellin moments, required to fix boundary conditions, are know.

\section{Conclusions}
In this paper we proposed a method for calculating phase-space integrals for the decay process \mbox{$1 \to n$} massless partons in QCD using integration-by-parts and differential equations techniques.
The key idea of our approach is based on choosing a basis of master integrals which leads to significant simplification of differential equations.
As a main result of this work, we describe an algorithm how to construct such a basis and find a solution of the resulting differential equations.
The advantage of our approach comparing to available techniques is that it is relatively simple to automate for execution on a computer without loss of generality of the final solution, which is obtained to any order in the dimensional regulator $\eps$ in terms of (generalized) harmonic polylogarithms.
That requires however to know at least one Mellin moment for every master integral in order to determine boundary conditions for the final solution.

In order to demonstrate how our method works in practice, we calculate master integrals for the decay processes $1 \to 4$ and $1 \to 3$ with a projection to $x$-space, needed to extract NLO time-like splitting functions from $e^+e^-$ annihilation process.
Analyzing this example we notice that another asset of the proposed method is that resulting master integrals are explicitly regulated in the singular points with the help of the dimensional regulator $\eps$, manifested by overall factors  $x^{-1+a\eps}$ and  $(1-x)^{-1+b\eps}$ in the final result.

The generalization of the results to NNLO topologies with loop insertions, needed to obtain missing $n_f^2$ pieces of the off-diagonal time-like splitting functions, is particularly straight-forward due to the factorizability of the phase-space.
In addition, master integrals with various types of projectors, not only to $x$-space as in the case of splitting functions, can be obtained with the described method as well.

\acknowledgments

I gratefully acknowledge the hospitality of the Theory Group of the University of Hamburg where major part of this research was done.
In particular, I am thankful for numerous discussions with prof. Sven-Olaf Moch, his support and guidance.
I also acknowledged useful discussions and comments from Johannes Henn and Roman Lee.

The Feynman diagrams were drawn with the help of \texttt{JaxoDraw}~\cite{BCKT08} and \texttt{Axodraw}~\cite{Ver94}.

This work has been supported by the Research Executive Agency (REA) with the European Union grant PITN-GA-2010-264564 (LHCPhenoNet) and by Narodowe Centrum Nauki with the Sonata Bis grant DEC-2013/10/E/ST2/00656.

\newpage \appendix

\section{NLO master integrals}

\subsection*{Real-virtual case}
\label{app:a}

Mellin moments of the real-virtual masters used to fix boundary conditions read \cite{MM06}
\begin{flalign}
  V_1(0) & = \frac{1}{8\eps} + \frac{5}{4} + \eps \left(8 - \frac{21}{16}\zeta_2\right),
  \\
  V_2(0) & = \frac{1}{8\eps} + \frac{17}{16} + \eps \left(\frac{183}{32} - \frac{17}{16}\zeta_2\right),
  \\
  V_3(0) & = \frac{1}{8\eps} + \frac{5}{4} + \eps \left(8 - \frac{21}{16}\zeta_2\right),
  \\
  V_4(0) & = -\frac{1}{4\eps} - \left(\frac{11}{4} - \frac{\zeta_2}{2}\right) + \eps \left(\frac{77}{4}+\frac{49}{8}\zeta_2+\frac{5}{2}\zeta_3\right),
  \\
  V_5(0) & = \frac{1}{8\eps^4} + \frac{1}{4\eps^3} + \frac{1}{\eps^2}\left(\frac{1}{2} - \frac{25}{16}\zeta_2\right),
  \\
  V_6(1) & = \frac{3}{8\eps^4} + \frac{3}{4\eps^3} + \frac{1}{\eps^2}\left(\frac{3}{2}-\frac{79}{16}\zeta_2\right).
  &&
\end{flalign}

Real-virtual master integrals read
\begin{flalign}
  V_1(x,\eps) & =
    \frac{x^{1-3\eps}(1-x)^{-\eps}}{8\eps}
    \Big\{
       2
     + 14 \eps
     + \eps^2 (66 -15\zeta_2)
    \Big\},
  \\
  V_2(x,\eps) & =
    \frac{x^{1-2\eps}(1-x)^{-\eps}}{48\eps}
    \Big\{
       12
     + 72 \eps
     + \eps^2 (288-13\zeta_2)
    \Big\},
  \\
  V_3(x,\eps) & =
    \frac{x^{1-2\eps}(1-x)^{-2\eps}}{48\eps}
    \Big\{
       12
     + 72 \eps
     + \eps^2 (288 - 13\zeta_2)
    \Big\},
  \\
  V_4(x,\eps) & =
    \frac{(1-x)^{-2\eps}}{4\eps}
    \Big\{
       \HPL_{1,0}
     + \eps \, (-\HPL_{1,2} - \HPL_{1,1,0} - 5\HPL_{1,0,0} + 2\HPL_{1,0}  - \zeta_2\HPL_1)
    \Big\},
  \\
  V_5(x,\eps) & =
    \frac{x^{-1-4\eps}}{4\eps^3}
    \Big\{
       -2
     - \eps \, (2\HPL_{0} + 4)
     + \eps^2 (-2\HPL_{1,1} - 2\HPL_{1,0} - 4\HPL_{1} - 17\zeta_2 - 8)
    \Big\},\!
  \\
  V_6(x,\eps) & =
    \frac{x^{-1-2\eps}(1-x)^{-1-2\eps}}{4\eps^3}
    \Big\{
       -3
     + \eps \, (3\HPL_{0}-6)
     - \eps^2 (3\HPL_{2} + 2\HPL_{1,0} + 3\HPL_{0,0} - 6\HPL_0
     \nonumber \\ & \quad
       - \frac{45}{2}\zeta_2 + 12)
    \Big\}.
  &&
\end{flalign}

\newpage
\subsection*{Real-real case}
\label{app:b}

Mellin moments of the real-real masters used to fix boundary conditions read \cite{MM06,GGH03}
\begin{flalign}
  R_1(0) & = \frac{1}{96} + \frac{71}{576}\eps + \eps^2 \left( \frac{3115}{3456} - \frac{7}{64}\zeta_2 \right),\!
  &
  R_2(0) & = \frac{1}{32} + \frac{25}{64}\eps + \eps^2 \left( \frac{383}{128} - \frac{21}{64}\zeta_2 \right),
  \\
  R_3(0) & = -\frac{1}{8\eps} - \frac{11}{8} - \eps \left( \frac{77}{8} - \frac{21}{16}\zeta_2\right),\!
  &
  R_4(0) & = \frac{5}{8\eps^4} + \frac{5}{4\eps^3} + \frac{1}{\eps^2}\left(\frac{5}{2}-\frac{105}{16}\zeta_2\right),
  \\
  R_5(0) & = -\frac{1}{16\eps^2} - \frac{3}{8\eps} - \left( \frac{13}{8} - \frac{17}{32}\zeta_2 \right),
  &
  R_6(0) & = \frac{\zeta_2-1}{8} + \eps \left(-\frac{7}{4} + \frac{7}{8}\zeta_2 + \frac{9}{8}\zeta_3\right),\!\!
  \\
  R_7(0) & = \frac{3}{32\eps^4} + \frac{3}{16\eps^3} + \frac{1}{\eps^2} \left( \frac{3}{8} - \frac{83}{64}\zeta_2\right),\!
  &
  R_7(1) & = \frac{1}{16\eps^4} + \frac{1}{8\eps^3} + \frac{1}{\eps^2} \left( \frac{1}{4} - \frac{29}{32}\zeta_2\right),
\end{flalign}
\vspace{-7mm}
\begin{flalign}
  R_8(0) & = \frac{3}{32\eps^4} + \frac{3}{16\eps^3} + \frac{1}{\eps^2} \left( \frac{3}{8} - \frac{83}{64}\zeta_2\right),
  \\
  R_8(1) & = \frac{\zeta_2}{4\eps^2} + \frac{1}{\eps}\left(\frac{19}{8}\zeta_3 + \frac{1}{2}\zeta_2\right) + \left( \frac{19}{4}\zeta_3 + \frac{133}{40}\zeta_2^2 + \zeta_2\right).
  &&
\end{flalign}

Real-real masters integrals read
\begin{flalign}
  R_1(x,\eps) & =
    \frac{x^{1-2\eps}(1-x)^{1-2\eps}}{64}
    \Big\{
       4
     + 34 \eps
     + \eps^2 (183 - 26 \zeta_2)
    \Big\},
  \\
  R_2(x,\eps) & =
    \frac{x^{1-3\eps}}{8}
    \Big\{
     - \HPL_{0}
     + \eps \left(-2\HPL_{2} - \HPL_{0,0} - 7\HPL_{0} + 2\zeta_2 \right)
     + \eps^2 \big( -2\HPL_{3} - 4\HPL_{2,1} - 2\HPL_{2,0} \nonumber \\ & \quad - \HPL_{0,0,0} - 14\HPL_{2} - 7\HPL_{0,0} + (-33+\frac{13}{2}\zeta_2)\HPL_{0} + 2\zeta_3 + 14\zeta_2 \big)
    \Big\},
  \\
  R_3(x,\eps) & =
    \frac{x^{-2\eps}}{8 \eps}
    \Big\{
       \HPL_{1,0}
     + \eps \left(-2\HPL_{1,1,0} - 5\HPL_{1,0,0} + 2\HPL_{1,0} + 2\zeta_2 \right)
    \Big\},
  \\
  R_4(x,\eps) & =
    \frac{x^{-1-2\eps} (1-x)^{-1-2\eps}}{16\eps^3}
    \Big\{
     - 10
     - 20 \eps
     - \eps^2 (40 - 65\zeta_2)
    \Big\},
  \\
  R_5(x,\eps) & =
    \frac{x^{-4\eps} (1-x)^{-2\eps} (1+x)^{-4\eps}}{16\eps^2}
    \Big\{
     - 1
     + \eps \left( - 2\HPL_{0} + 4\HPL_{-1} - 2 \right)
     + \eps^2 \big( - 2\HPL_{0,0}
     \nonumber \\ & \quad
       + 8\HPL_{-1,0} - 16\HPL_{-1,-1} + 8\HPL_{-2} - 4\HPL_{0} + 8\HPL_{-1} + \frac{9}{2}\zeta_2 - 4 \big)
    \Big\},
  \\
  R_6(x,\eps) & =
    \frac{(1+x)^{-6\eps}}{4}
    \Big\{
       x \HPL_{0,0}
     - (1+x) \HPL_{-1,0}
     - x \, \zeta_2
    \Big\},
  \\
  R_7(x,\eps) & =
    \frac{x^{-1-2\eps} (1-x)^{-1-2\eps}}{8\eps^3}
    \Big\{
     - 1
     + \eps \left( 2 \HPL_{0} - 2 \right)
     + \eps^2 \big( -2 \HPL_{1,0} - 4 \HPL_{0,0} + 4 \HPL_{0}
   \nonumber \\ & \quad
     + \frac{9}{2}\zeta_2 - 4 \big)
    \Big\},
  \\
   R_8(x,\eps) & =
     \frac{x^{-1-2\eps} (1-x)^{-1-2\eps}}{4\eps^2}
     \Big\{
      - 2 \HPL_{0}
      + \eps ( - 4\HPL_{2} + \HPL_{0,0} - 2\HPL_{-1,0} - 4\HPL_{-2} - 4\HPL_{0}
   \nonumber \\ & \quad
      + 5\zeta_2 ) + \eps^2 \big( 2\HPL_{3} - 8\HPL_{2,1} - 8\HPL_{2,-1} + \HPL_{0,0,0} - 4\HPL_{-1,2} - 4\HPL_{-1,-2}
   \nonumber \\ & \quad
      - 8\HPL_{-2,1} - 8\HPL_{-2,-1} + 2\HPL_{-3} - 8\HPL_{2} + 2\HPL_{0,0} - 4\HPL_{-1,0} - 8\HPL_{-2}
   \nonumber \\ & \quad
      + (9\zeta_2 - 8)\HPL_0 + 6\zeta_2\HPL_{-1} + 7\zeta_3 + 10\zeta_2 \big)
     \Big\}.
  &&
\end{flalign}

\newpage
\bibliography{bibliography} 

\providecommand{\href}[2]{#2}\begingroup\raggedright\begin{thebibliography}{10}

\bibitem{FGK89}
N.~K. Falck, D.~Graudenz, and G.~Kramer, {\it {Cross-section for Five Jet
  Production in $e^+ e^-$ Annihilation}},  {\em Nucl. Phys.} {\bf B328} (1989)
  317.

\bibitem{BDK97}
Z.~Bern, L.~J. Dixon, and D.~A. Kosower, {\it {One loop amplitudes for e+ e- to
  four partons}},  {\em Nucl. Phys.} {\bf B513} (1998) 3--86,
  [\href{http://arxiv.org/abs/hep-ph/9708239}{{\tt hep-ph/9708239}}].

\bibitem{GGGKR01}
L.~Garland, T.~Gehrmann, E.~N. Glover, A.~Koukoutsakis, and E.~Remiddi, {\it
  {The Two loop QCD matrix element for e+ e- ---> 3 jets}},  {\em Nucl.Phys.}
  {\bf B627} (2002) 107--188, [\href{http://arxiv.org/abs/hep-ph/0112081}{{\tt
  hep-ph/0112081}}].

\bibitem{MUW02}
S.~Moch, P.~Uwer, and S.~Weinzierl, {\it {Two loop amplitudes with nested sums:
  Fermionic contributions to e+ e- ---> q anti-q g}},  {\em Phys. Rev.} {\bf
  D66} (2002) 114001, [\href{http://arxiv.org/abs/hep-ph/0207043}{{\tt
  hep-ph/0207043}}].

\bibitem{MVV04a}
S.~Moch, J.~Vermaseren, and A.~Vogt, {\it {The Three loop splitting functions
  in QCD: The Nonsinglet case}},  {\em Nucl.Phys.} {\bf B688} (2004) 101--134,
  [\href{http://arxiv.org/abs/hep-ph/0403192}{{\tt hep-ph/0403192}}].

\bibitem{MVV04b}
A.~Vogt, S.~Moch, and J.~Vermaseren, {\it {The Three-loop splitting functions
  in QCD: The Singlet case}},  {\em Nucl.Phys.} {\bf B691} (2004) 129--181,
  [\href{http://arxiv.org/abs/hep-ph/0404111}{{\tt hep-ph/0404111}}].

\bibitem{MMV06}
A.~Mitov, S.~Moch, and A.~Vogt, {\it {Next-to-Next-to-Leading Order Evolution
  of Non-Singlet Fragmentation Functions}},  {\em Phys.Lett.} {\bf B638} (2006)
  61--67, [\href{http://arxiv.org/abs/hep-ph/0604053}{{\tt hep-ph/0604053}}].

\bibitem{MV08}
S.~Moch and A.~Vogt, {\it {On third-order timelike splitting functions and
  top-mediated Higgs decay into hadrons}},  {\em Phys.Lett.} {\bf B659} (2008)
  290--296, [\href{http://arxiv.org/abs/0709.3899}{{\tt arXiv:0709.3899}}].

\bibitem{AMV11}
A.~Almasy, S.~Moch, and A.~Vogt, {\it {On the Next-to-Next-to-Leading Order
  Evolution of Flavour-Singlet Fragmentation Functions}},  {\em Nucl.Phys.}
  {\bf B854} (2012) 133--152, [\href{http://arxiv.org/abs/1107.2263}{{\tt
  arXiv:1107.2263}}].

\bibitem{ARS15}
D.~P. Anderle, F.~Ringer, and M.~Stratmann, {\it {Fragmentation Functions at
  Next-to-Next-to-Leading Order Accuracy}},
  \href{http://arxiv.org/abs/1510.05845}{{\tt arXiv:1510.05845}}.

\bibitem{GGG05}
A.~Gehrmann-De~Ridder, T.~Gehrmann, and E.~W.~N. Glover, {\it {Antenna
  subtraction at NNLO}},  {\em JHEP} {\bf 09} (2005) 056,
  [\href{http://arxiv.org/abs/hep-ph/0505111}{{\tt hep-ph/0505111}}].

\bibitem{STD06}
G.~Somogyi, Z.~Trocsanyi, and V.~Del~Duca, {\it {A Subtraction scheme for
  computing QCD jet cross sections at NNLO: Regularization of doubly-real
  emissions}},  {\em JHEP} {\bf 01} (2007) 070,
  [\href{http://arxiv.org/abs/hep-ph/0609042}{{\tt hep-ph/0609042}}].

\bibitem{Cza10}
M.~Czakon, {\it {A novel subtraction scheme for double-real radiation at
  NNLO}},  {\em Phys. Lett.} {\bf B693} (2010) 259--268,
  [\href{http://arxiv.org/abs/1005.0274}{{\tt arXiv:1005.0274}}].

\bibitem{CT81}
K.~Chetyrkin and F.~Tkachov, {\it {Integration by Parts: The Algorithm to
  Calculate beta Functions in 4 Loops}},  {\em Nucl.Phys.} {\bf B192} (1981)
  159--204.

\bibitem{Tka81}
F.~Tkachov, {\it {A Theorem on Analytical Calculability of Four Loop
  Renormalization Group Functions}},  {\em Phys.Lett.} {\bf B100} (1981)
  65--68.

\bibitem{Smi04}
V.~Smirnov, {\it {Evaluating Feynman integrals}},  {\em Springer Tracts
  Mod.Phys.} {\bf 211} (2004) 1--244.

\bibitem{GMTW14}
T.~Gehrmann, A.~von Manteuffel, L.~Tancredi, and E.~Weihs, {\it {The two-loop
  master integrals for $q\overline{q} \to VV$}},  {\em JHEP} {\bf 1406} (2014)
  032, [\href{http://arxiv.org/abs/1404.4853}{{\tt arXiv:1404.4853}}].

\bibitem{MSZ14}
A.~von Manteuffel, R.~M. Schabinger, and H.~X. Zhu, {\it {The two-loop soft
  function for heavy quark pair production at future linear colliders}},  {\em
  Phys. Rev.} {\bf D92} (2015), no.~4 045034,
  [\href{http://arxiv.org/abs/1408.5134}{{\tt arXiv:1408.5134}}].

\bibitem{ADDHM15}
C.~Anastasiou, C.~Duhr, F.~Dulat, F.~Herzog, and B.~Mistlberger, {\it {Higgs
  Boson Gluon-Fusion Production in QCD at Three Loops}},  {\em Phys. Rev.
  Lett.} {\bf 114} (2015) 212001, [\href{http://arxiv.org/abs/1503.06056}{{\tt
  arXiv:1503.06056}}].

\bibitem{AHHHKS15}
C.~Anzai, A.~Hasselhuhn, M.~Höschele, J.~Hoff, W.~Kilgore, M.~Steinhauser, and
  T.~Ueda, {\it {Exact N$^{3}$LO results for qq$^{′}$ → H + X}},  {\em
  JHEP} {\bf 07} (2015) 140, [\href{http://arxiv.org/abs/1506.02674}{{\tt
  arXiv:1506.02674}}].

\bibitem{BBDMS15}
A.~Behring, J.~Blümlein, A.~De~Freitas, A.~von Manteuffel, and C.~Schneider,
  {\it {The 3-Loop Non-Singlet Heavy Flavor Contributions to the Structure
  Function $g_{1}(x,Q^{2})$ at Large Momentum Transfer}},  {\em Nucl. Phys.}
  {\bf B897} (2015) 612--644, [\href{http://arxiv.org/abs/1504.08217}{{\tt
  arXiv:1504.08217}}].

\bibitem{RN96}
P.~Rijken and W.~van Neerven, {\it {Higher order QCD corrections to the
  transverse and longitudinal fragmentation functions in electron - positron
  annihilation}},  {\em Nucl.Phys.} {\bf B487} (1997) 233--282,
  [\href{http://arxiv.org/abs/hep-ph/9609377}{{\tt hep-ph/9609377}}].

\bibitem{GGH03}
A.~Gehrmann-De~Ridder, T.~Gehrmann, and G.~Heinrich, {\it {Four particle phase
  space integrals in massless QCD}},  {\em Nucl.Phys.} {\bf B682} (2004)
  265--288, [\href{http://arxiv.org/abs/hep-ph/0311276}{{\tt hep-ph/0311276}}].

\bibitem{MV99}
S.~Moch and J.~Vermaseren, {\it {Deep inelastic structure functions at two
  loops}},  {\em Nucl.Phys.} {\bf B573} (2000) 853--907,
  [\href{http://arxiv.org/abs/hep-ph/9912355}{{\tt hep-ph/9912355}}].

\bibitem{MM06}
A.~Mitov and S.~Moch, {\it {QCD Corrections to Semi-Inclusive Hadron Production
  in Electron-Positron Annihilation at Two Loops}},  {\em Nucl.Phys.} {\bf
  B751} (2006) 18--52, [\href{http://arxiv.org/abs/hep-ph/0604160}{{\tt
  hep-ph/0604160}}].

\bibitem{Henn13}
J.~M. Henn, {\it {Multiloop integrals in dimensional regularization made
  simple}},  {\em Phys. Rev. Lett.} {\bf 110} (2013) 251601,
  [\href{http://arxiv.org/abs/1304.1806}{{\tt arXiv:1304.1806}}].

\bibitem{Mos60}
J.~Moser, {\it The order of a singularity in fuchs' theory},  {\em
  Mathematische Zeitschrift} {\bf 72} (1959), no.~1 379--398.

\bibitem{Henn14}
J.~M. Henn, {\it {Lectures on differential equations for Feynman integrals}},
  {\em J. Phys.} {\bf A48} (2015) 153001,
  [\href{http://arxiv.org/abs/1412.2296}{{\tt arXiv:1412.2296}}].

\bibitem{Lee14}
R.~N. Lee, {\it {Reducing differential equations for multiloop master
  integrals}},  {\em JHEP} {\bf 04} (2015) 108,
  [\href{http://arxiv.org/abs/1411.0911}{{\tt arXiv:1411.0911}}].

\bibitem{GM15}
O.~Gituliar and S.~Moch, {\it {Towards three-loop QCD corrections to the
  time-like splitting functions}},  {\em Acta Phys. Polon.} {\bf B46} (2015),
  no.~7 1279--1289, [\href{http://arxiv.org/abs/1505.02901}{{\tt
  arXiv:1505.02901}}].

\bibitem{NW93}
P.~Nason and B.~Webber, {\it {Scaling violation in e+ e- fragmentation
  functions: QCD evolution, hadronization and heavy quark mass effects}},  {\em
  Nucl. Phys.} {\bf B421} (1994) 473--517. [Erratum: Nucl.
  Phys.B480,755(1996)].

\bibitem{Nog91}
P.~Nogueira, {\it {Automatic Feynman graph generation}},  {\em J.Comput.Phys.}
  {\bf 105} (1993) 279--289.

\bibitem{Ver00}
J.~{Vermaseren}, {\it {New features of FORM}},
  \href{http://arxiv.org/abs/math-ph/0010025}{{\tt math-ph/0010025}}.

\bibitem{Lee12}
R.~Lee, {\it {Presenting LiteRed: a tool for the Loop InTEgrals REDuction}},
  \href{http://arxiv.org/abs/1212.2685}{{\tt arXiv:1212.2685}}.

\bibitem{Lee13}
R.~Lee, {\it {LiteRed 1.4: a powerful tool for reduction of multiloop
  integrals}},  {\em J.Phys.Conf.Ser.} {\bf 523} (2014) 012059,
  [\href{http://arxiv.org/abs/1310.1145}{{\tt arXiv:1310.1145}}].

\bibitem{RV99}
E.~Remiddi and J.~Vermaseren, {\it {Harmonic polylogarithms}},  {\em Int. J.
  Mod. Phys.} {\bf A15} (2000) 725--754,
  [\href{http://arxiv.org/abs/hep-ph/9905237}{{\tt hep-ph/9905237}}].

\bibitem{AB04}
U.~Aglietti and R.~Bonciani, {\it {Master integrals with 2 and 3 massive
  propagators for the 2 loop electroweak form-factor - planar case}},  {\em
  Nucl. Phys.} {\bf B698} (2004) 277--318,
  [\href{http://arxiv.org/abs/hep-ph/0401193}{{\tt hep-ph/0401193}}].

\bibitem{BDV10}
R.~Bonciani, G.~Degrassi, and A.~Vicini, {\it {On the Generalized Harmonic
  Polylogarithms of One Complex Variable}},  {\em Comput. Phys. Commun.} {\bf
  182} (2011) 1253--1264, [\href{http://arxiv.org/abs/1007.1891}{{\tt
  arXiv:1007.1891}}].

\bibitem{ABS13}
J.~{Ablinger}, J.~{Bl{\"u}mlein}, and C.~{Schneider}, {\it {Analytic and
  algorithmic aspects of generalized harmonic sums and polylogarithms}},  {\em
  Journal of Mathematical Physics} {\bf 54} (Aug., 2013) 082301,
  [\href{http://arxiv.org/abs/1302.0378}{{\tt arXiv:1302.0378}}].

\bibitem{Maitre05}
D.~Maitre, {\it {HPL, a mathematica implementation of the harmonic
  polylogarithms}},  {\em Comput. Phys. Commun.} {\bf 174} (2006) 222--240,
  [\href{http://arxiv.org/abs/hep-ph/0507152}{{\tt hep-ph/0507152}}].

\bibitem{Smi08}
A.~Smirnov, {\it {Algorithm FIRE -- Feynman Integral REduction}},  {\em JHEP}
  {\bf 0810} (2008) 107, [\href{http://arxiv.org/abs/0807.3243}{{\tt
  arXiv:0807.3243}}].

\bibitem{MS12}
A.~von Manteuffel and C.~Studerus, {\it {Reduze 2 - Distributed Feynman
  Integral Reduction}},  \href{http://arxiv.org/abs/1201.4330}{{\tt
  arXiv:1201.4330}}.

\bibitem{BCKT08}
D.~Binosi, J.~Collins, C.~Kaufhold, and L.~Theussl, {\it {JaxoDraw: A Graphical
  user interface for drawing Feynman diagrams. Version 2.0 release notes}},
  {\em Comput. Phys. Commun.} {\bf 180} (2009) 1709--1715,
  [\href{http://arxiv.org/abs/0811.4113}{{\tt arXiv:0811.4113}}].

\bibitem{Ver94}
J.~A.~M. Vermaseren, {\it {Axodraw}},  {\em Comput. Phys. Commun.} {\bf 83}
  (1994) 45--58.

\end{thebibliography}\endgroup

\end{document}